\documentclass[epj]{svjour}

\usepackage{latexsym}
\usepackage{graphicx}
\newcommand{\be}[1]{\begin{equation}\label{#1}}
\newcommand{\ee}{\end{equation}}     
\newcommand{\bea}{\begin{eqnarray}}
\newcommand{\eea}{\end{eqnarray}} 
\newcommand{\eq}[1]{Eq.\,(\ref{#1})}
\newcommand{\fig}[1]{Fig.\,\ref{#1}}
\begin{document}
\title{Electron dynamics in strong laser pulse illumination \\
of large rare gas clusters} 
\author{Ulf Saalmann and Jan M.~Rost
}                     
%
%
\institute{Max Planck Institute for 
the Physics of Complex Systems, N\"othnitzer 
 Str.~38, 01187 Dresden, Germany}
\date{Received: date / Revised version: date}
%
\abstract{
We analyze the dynamics of up to $10^5$ electrons resulting
from illuminating a Xenon cluster with 9093 atoms with intense laser
pulses of different length and peak intensity. Interesting details of
electron motion are identified which can be probed with a time
resolution of 100 attoseconds. 
Corresponding experiments would shed light on unexplored
territory in complex electronic systems such as clusters and they
would also permit to critically access the present theoretical
description of this dynamics.  
} 
\maketitle
\section{Introduction}
Recent advance in laser technology has led to the creation of
subfemtosecond (100 attosecond) strong laser pulses which allow one in
the future to resolve much shorter time scales in microscopic dynamics
\cite{heki+01,nile+02} than it has been possible so far. More
specifically, while until know vibrational motion in molecules could
be  resolved in time experimentally, 
it will be possible to follow electronic motion with attosecond resolution. 
Still, the atomic time unit, i.e., the period of the ground state
electron in hydrogen, is of the order of $10^{-17}$\,s, i.e., roughly
10 attoseconds while for now a pulse length of 100 attoseconds and
more is experimentally feasible.
However, there are alternative methods such as tomographic imaging
which use the high harmonic spectrum of a femtosecond pulse to get
attosecond resolved dynamics. In this way it could be demonstrated
that the {\it electronic} wavefunction of a molecule could be directly
imaged \cite{itle+04}. This is fantastic progress, yet one might argue
that we know in principle the electronic wavefunction of a diatomic
molecule. We can even calculate it, and as it turns 
out experimental imaging and theoretical calculations via a standard
solution of the Schr\"odinger equation agree well \cite{itle+04}. 

On the other hand, for systems more complex than a diatomic molecule,
e.g., a cluster consisting of many atoms, we neither know with
certainty theoretically the dynamics of the electrons (we have to make
model assumptions) nor do we have a way so far to access the electron
dynamics time resolved in the experiment. 

Pioneering experiments 
\cite{kosc+99dofe+05} 
have demonstrated non-tri\-vial time dependent dynamics in
cluster motion under a strong laser 
pulse. This interesting dynamics can be attributed to resonance like
behavior where either the next neighbor ions contribute in a
cooperative way to enhance ionization \cite{siro02} or the electrons,
still bound to the cluster but not to individual atoms, enter a
collective phase of motion  which is susceptible to resonant
absorption of radiation \cite{saro03}. 
Besides that, a nonlinear excitation of this resonance
could be relevant \cite{fopo+03}.
A slightly different situation results from irradiation with VUV
pulses as realized with the free electron laser \cite{wabi+02}. 
There, inverse Bremsstrahlung seems to be
the main mechanism of energy absorption. Details, however, remain
controversial \cite{sagr03,siro04}. Detailed knowledge would require the
experimental ability to time resolve the electron motion. Since the
electrons are no longer strongly bound to a single ion but to the
extended cluster, there typical time scale of motion is a bit slower
which would be ideal for  probing it with subfemtosecond laser light.

It is the purpose of this paper to explore which kind of details the
electron dynamics exhibits during a standard femtosecond laser pulse  
a length of the order of 100\,fs. 
Specifically, we will concentrate on energy spectra of electrons from
a Xenon cluster with 9093 atoms during the irradiation with a laser
pulse of length from 25 to 400\,fs and peak intensity ranging from
$10^{14}$ to $10^{16}$\,W/cm$^2$.  

\section{Theoretical method}
Before we start to discuss the spectra we give a brief account of the
theoretical method how these spectra have been obtained. The initial
cluster configuration is derived assuming an icosahedral symmetry of
the atoms. The faces of the $n$th shell contain $(n{+}1)(n{+}2)/2$
atoms.   
Keeping only those atoms which are inside a sphere of radius
$R_0 = 50$\,{\AA} yields a cluster with 9093 atoms. 
This configuration is relaxed using pairwise Lennard-Jones
potentials to find the optimum interatomic distances thereby
freezing the cluster geometry.  
 
The laser pulse has the form of a Gaussian (we use atomic units if not
stated otherwise) 
\bea
\vec F(t) &=& \hat{\vec{z}} F_t\cos\omega t
\nonumber\\
&=& \hat{\vec{z}}
\sqrt{I}
\exp\left[\log2\,\left(2t/T\right)^2\right]
\cos\omega t 
\label{pulse} 
\eea
with half width $T$ which measures the pulse length, peak intensity
$I$ and the laser frequency of $\omega = 0.058$\,a.u.\ 
(corresponding to 780\,nm wavelength). 
The light is linearly polarized along the $z$-axis,
denoted by the unit vector $\hat{\vec{z}}$. 
To make the simulation tractable we only model explicitly the
outermost bound electron of an atom/ion \cite{saro03}. 
In this way more and more electrons ``are generated'' due to the
removal from their mother ion during the laser pulse. 
The mutual attractive Coulomb potentials have
soft cores. All existing charged particles, ions and electrons, are
propagated classically exposed to their mutual attraction and
repulsion as well as to the coupling with the light pulse. Handling up
to about ten electrons per atom implies a total of up to 10$^5$ particles
to be propagated over relative long times 
(of the order of hundred optical cycles). 
This is not possible with standard molecular dynamics but
requires tree-code techniques \cite{pfgi96}. 
Alternatively, one can start from the particle-in-cell concept
to handle clusters of this size \cite{juge+04}.

\section{Energy spectra of cluster electrons during the laser pulse}

In this section we present and interpret the energy spectra of the electrons.
An overview of the spectra is given in \fig{spectra}, the parameters
for each panel of figure 
\fig{spectra} are provided in Table~\ref{tab:1}. Basically, for each
pulse length $T$ the 
peak intensity increases by a factor of 5 from top to bottom
 (set A to C). 
For increasing pulse lengths $T$ from left to right the fluency of the
pulse remains the same. Hence, from one figure to the next the peak
intensity decreases by a factor of 2 since the pulse length increases
by a factor of 2. All figures show distributions of total
energies of the electrons on 
the $y$-axis ranging from $-10$\,keV to $+5$\,keV
with the total energy of electron $i$ at position  $\vec r_i$
with momentum $\vec p_i$ defined as
\be{eq:teei}
E_i=\frac{\vec p_i^2}{2}-\sum_a^\mathrm{ions}
\frac{q_a}{|\vec r_a{-}\vec r_i|}+
\!\!\sum_{j(\ne i)}^\mathrm{electrons}\!\!\frac{1}{|\vec r_j{-}\vec r_i|}
\ee
with $\vec r_a$ and $q_a$ the position and the charge of ion
$a$, respectively.

The time evolution of the spectrum is shown from $-T$ to
$\max\{+T,\,100\,\mbox{fs}\}$.  
Of course, the calculation of the dynamics sets in long before $-T$,
when the pulse intensity is still negligible. 
It should be pointed out that
\fig{spectra} 
contains for each plot and at each time the energies of up to
$10^5$ electrons. Just the generation of such a plot in a
reasonable time requires to use the tree-code information. 

\subsection{Common features in all spectra} 
\begin{figure*}
\centering
\includegraphics[angle=90]{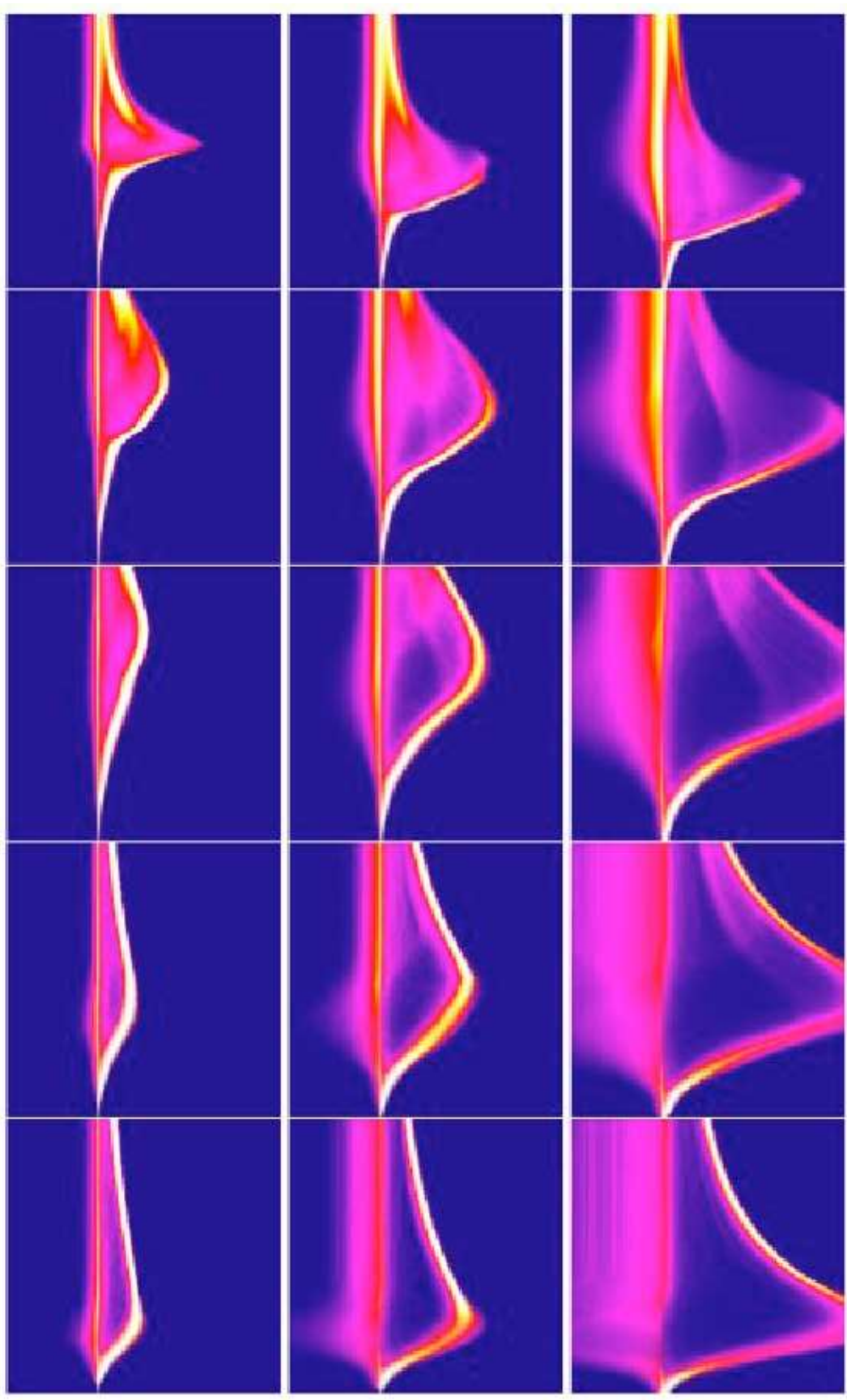}
\caption{Energy distribution of electrons 
according to \eq{eq:teei} under illumination of a
Xe$_{9093}$ cluster with laser pulses of different pulse length $T$
and peak intensity $I$ as detailed in Table~\ref{tab:1}. The pulse
shape is given in \eq{pulse}. The visible area 
has range of energy on the $y$-axis between $-10$\,keV and $+5$\,keV, and
a range of time on the $x$-axis between $-T$ and
$\max\{T,100\,\mathrm{fs}\}$.} 
\label{spectra}       
\end{figure*}

A first orientation in the plots reveals that the lowest electron
energies go always through a minimum after the laser pulse has reached
its maximum intensity. The minimum assumes quite different values, for
the parameters in \fig{spectra} roughly between $-2$ and $-15$\,keV. 
The negative electron energies are due to the positive background charge
$Q_t=\sum_a q_a$ of
the cluster which is a direct consequence of ionized electrons which
have left the cluster. They appear at positive energies in the
plots. As stronger the cluster is ionized as deeper the remaining
electrons are bound by the back ground charge. Upon explosion of the
ions the background charge spreads out and consequently, the energy of
the trapped electrons decreases for longer times. 
Evident are two preferred energy regions (with high intensities) for
the electrons: 
Ionized electrons have excess energies close to $E=0$. The
trapped electrons prefer energies close to the lowest possible
values. 

\begin{figure*}
\centering
\includegraphics[scale=0.9]{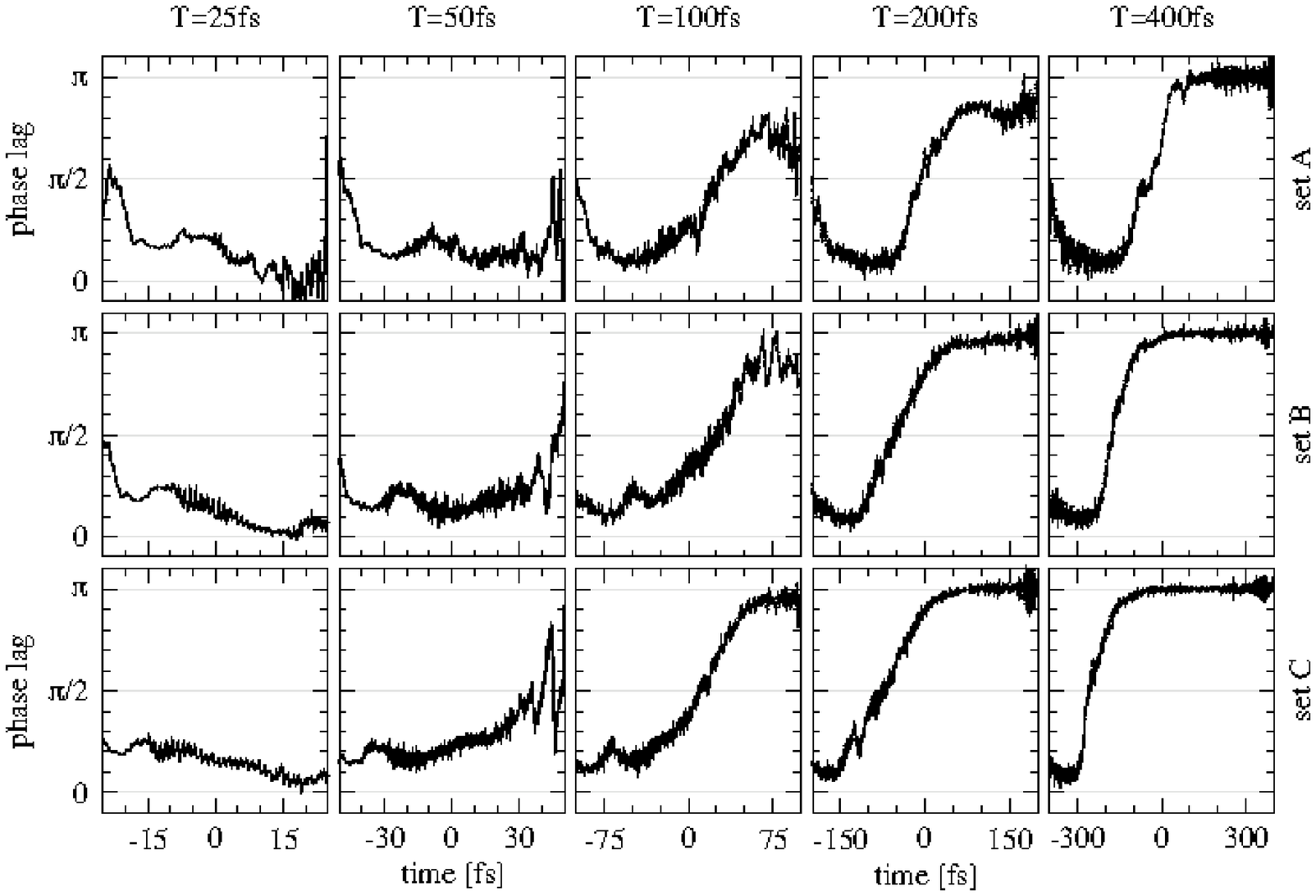}
\caption{Phase difference between laser cycle and periodic
  response of electrons inside the cluster volume for the
  same laser pulse parameters as shown in \fig{spectra}.}  
\label{phaselag}       
\end{figure*}

\subsection{General trends as a function of peak intensity: \\
The role of the quiver amplitude} 
The population of the two preferred regions is most prominent in the
lower left corner of \fig{spectra}, i.e., for maximum peak
intensity. For minimum peak intensity (upper right corner) 
the opposite trend can be identified: A substantial number of
electrons fills the area (of negative energy) between the two
preferred regions. 
If one compares the equilibrium size of the cluster given by the
radius $R_0 = 50$\,{\AA} with the 
quiver amplitude  of an electron in the laser field $x_\omega =
\sqrt{I}/\omega^2$, listed in Table~\ref{tab:1}, the reason for this
trend becomes obvious: 
For largest intensities (lower left corner) $x_\omega > R_0$ and 
the ionized electrons are driven far outside the cluster. The opposite
is true for low intensities 
(upper right corner) with the lowest intensities where despite
quivering in the laser field the electrons remain well inside the
cluster, leading to a distribution of negative electron energies due
to the attractive background charge.

\begin{table}
\caption{Laser pulse parameters for the 15 plots in \fig{spectra}.
The peak intensity $I$ given in units of 10$^{15}$W/cm$^2$ and the
pulse length $T$ in fs enters \eq{pulse}.
Furthermore, the quiver amplitude $x_\omega=\sqrt I/\omega^2$ is
given in {\AA} and the ponderomotive energy $U_\mathrm{p}=I/4\omega^2$
in units of eV, whereby $\omega = 0.058$\,a.u., the laser frequency
corresponding to 780\,nm wavelength.} 
\centering
\label{tab:1}       
\begin{tabular}{cc||r|r|r|r|r|}
& T & 25\,\, & 50\,\, & 100 & 200 & 400\\
\hline\hline 
set & $I$ & $3.2$ & $1.6$ & $0.8$ & $0.4$ & $0.2$  \\
A & $x_\omega$ & 46.7 & 33.1 & 23.4 & 16.5 & 11.7 \\
& $U_\mathrm{p}$ & 181 & 90.7 & 45.4 & 22.7 & 11.3 \\
\hline
set &$I$ & 16 & 8 & 4 & 2 & 1 \\
B & $x_\omega$  &  105 & 73.9 & 52.3 & 37.0 & 26.1 \\
& $U_\mathrm{p}$ & 907 & 454 & 227 & 113 & 56.7 \\
\hline
set &$I$ & 80 & 40 & 20 & 10 & 5 \\
C & $x_\omega$  & 234 & 165 & 117 & 82.6 & 58.4 \\
& $U_\mathrm{p}$ & 4536 & 2268 & 1134 & 567 & 284 \\
\hline
\end{tabular}
\end{table}

\subsection{The trace of escaping single ions}
In the panels corresponding to large pulse lengths and high peak
intensity (lower right corner of \fig{spectra}) one sees individual
lines of electron energies 
above the energy minimum and converging to each other for longer times.
These lines are energies of electron being trapped in excited orbits
around a single ion due to rising interatomic barriers when the
cluster expands. 
Hence, large pulse lengths are necessary for this feature. It also
occurs  in the upper right corner of \fig{spectra} (i.e., for lower
intensities) but it is masked there by other electrons with similar
energies. They cannot leave the cluster since the laser intensity is
too low.

\subsection{Atomic versus cluster effects}

Very short pulses in small clusters explore primarily atomic
properties of the atoms and ions within the cluster since there is not
enough time for a cooperative response of the cluster as a whole, e.g,
by expansion or thermalized collective electron motion. Large
clusters, however, develop upon the first atomic field ionization for
each atom a substantial positive background charge (in our case 
roughly $Q=+10^4$ for single ionization of the atoms). The main
mechanism for further ionization is therefore the escape from the
potential generated by the background charge. It can be modelled by
assuming a homogeneously charged sphere with total charge $Q_t$ and
(cluster) radius $R_t$, which leads to the potential 
\be{eq:clpo}
V_t(r)=\left\{
  \begin{array}{ll}
    -Q_t(3R_t{}^2{-}r^2)/(2R_t{}^3) & \quad\mbox{if } r\le R_t\\ [1ex]
    -Q_t/r  & \quad\mbox{if } r\ge R_t\,,
  \end{array}
\right.
\end{equation}
where quantities with subscript $t$  are weakly time dependent. 
All electrons in this potential can leave the cluster if the height of
the potential barrier at maximum field strength within a laser
period is lower or equal to the potential minimum. 
This happens at the field strength $F_t$ that shifts the potential
minimum from $r=0$ to the cluster surface $r=R_t$, i.e.\ 
$\left[dV_t(r)/dr\right]_{R_t}=F_t$.
From this the maximum charging of the cluster 
can be estimated to be
\be{charging}
Q_t = R_t^2 F_t\,.
\ee
The depth of the minima in the  energy of the electrons in \fig{spectra}
indicates directly the charging of the cluster. Hence, if \eq{charging} is true
the depth and the early charging process in time should agree if the energy
($y$-axis) in the panels is scaled by $F_t$. Indeed, for the three
panels A--C belonging to $T=25$\,fs and $T=50$\,fs the temporal development
of the electronic energy until the minimum is reached agrees very well
after scaling (not shown). The considerations apply as long as the
cluster has not expanded yet, i.e., $R_t\approx R_0$. For longer times the
cluster expands and the scaling does not apply.

\subsection{Resonant energy absorption through collective electron motion}

If the pulse is long enough to allow for an expansion of the cluster
during the illumination by laser light resonant absorption becomes
possible. The mechanism behind resonant absorption in a cluster of the
present size is the match between the frequency of collective electron
motion  within the cluster and the laser frequency. This leads to a
characteristic phase lag of $\pi/2$ between the oscillation of the
laser amplitude and the electronic response \cite{saro03}. 
If one compares \fig{phaselag} with \fig{spectra} one sees that
resonant absorption  
goes hand in gloves with a sharp decrease  of the minimum electron
energy (i.e. a fast charging of the cluster).  This is clearly visible
in all panels for $T = 400$\,fs and in panel $A$ for $T=200$\,fs. If
the resonance occurs during the fall of the laser envelope ($t>0$) it
cannot be directly recognized in the electron energy (compare
\fig{spectra} with \fig{phaselag}).

\subsection{Fine structure on the scale of the optical cycle}
\begin{figure}
  \centering
  \includegraphics[scale=0.6]{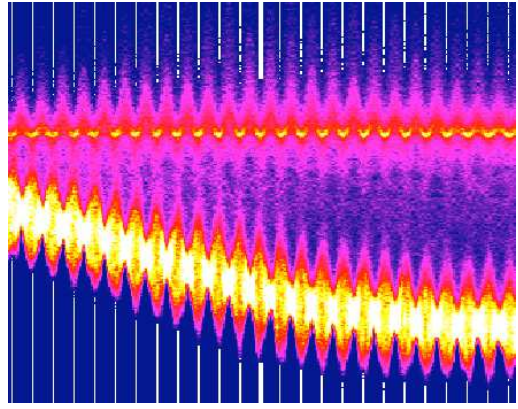}
  \caption{Magnification of the electron spectrum of series A at
    $T=50$\,fs from \fig{spectra}. The white lines indicate times of maximum
    field strength, the thick white line indicates the time of the cut
    through the spectrum shown in \fig{cut}.} 
  \label{spectra_A50}  
\end{figure}

\begin{figure}
\centering
   \includegraphics[scale=0.5]{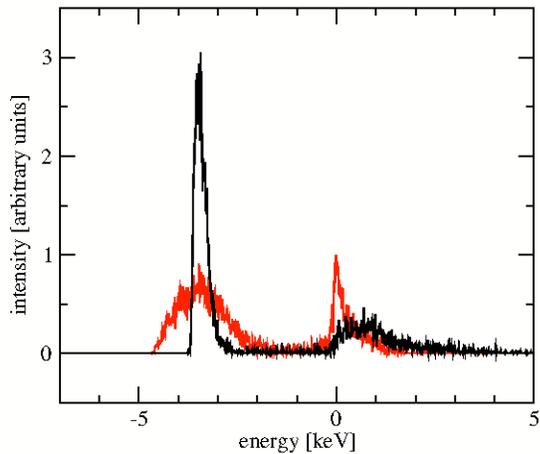}
\caption{Cut through the electron energy spectrum of series A at 50\,fs
in the optical cycle around maximum 
laser intensity $I$ (time $t=0$) indicated by a thick white line in
\fig{spectra_A50}. The black curve is for almost vanishing field
strength, the red (grey) curve for the maximal field strength
within the optical cycle.} 
 \label{cut}  
 \end{figure}

Finally, we would like to draw attention to a peculiar behavior in the 
periodic response of continuum electrons and bound electrons to the
driving laser. 
Most intensity of such electrons can be found close to E = 0 and close
to the minimum of the electron energies, respectively. Firstly one
sees that
the spots of high electron density occur for bound and continuum electrons
with a phase difference of half a period, cf.\ \fig{spectra_A50}. Secondly,
and may be even more puzzling on a first glance, the energy
distribution of the electrons seems to ``breeze'', with a narrow
distribution at zero field and a wide distribution at maximum field
for the bound electrons and exactly the opposite, although less
pronounced, for the continuum electrons, cf.\ \fig{cut}. 

The behavior of the continuum electrons ($E_i>0$)
in \fig{spectra_A50} is easily understood. 
They have a certain spread  of drift velocities $\overline{\vec{p}}_i$
or momenta as a result of the energy absorption and the
individual electron has an energy of 
$E_i = (\overline{\vec{p}}_i+\vec A(t))^2/2$ with 
$\vec A(t) = \hat{\vec{z}}\int^t \mathrm{d}\tau\, F(\tau) 
\approx \hat{\vec{z}} F_t/\omega\sin\omega t
= \hat{\vec{z}} F_t/\omega\cos(\omega t-\pi/2)$. 
In other words, since these electrons oscillate with a phase lag
of $\pi$ the velocity or the momentum has a phase lag of $\pi/2$.
Hence, the spread of the $\overline{\vec{p}}_i$ 
is amplified due to the squaring in the energy $E_i$ at zero
field, as can be seen from the peaks between the white lines in
the upper part of \fig{spectra_A50} and from the corresponding
distribution of energies $E_i>0$ in \fig{cut} (black curve).

The energies $E_i$ of the bound electrons ($E_i<0$ in 
\fig{spectra_A50}) show a similar spreading 
phenomenon, however with maximum spread at maximum field. This can be
understood if one takes into consideration that the many electrons in
the cluster are an interacting system of charges which reacts very
fast to forces, trying to reach an equilibrium position and
thereby maintaining the spatial distribution in the cluster.
In a first approximation it can be assumed to be incompressible.
This distribution is driven by the laser and has a vanishing
phase lag since the eigenfrequency $\Omega_t^2=Q_t/R_t^3$
of the cluster potential
\eq{eq:clpo} is larger than the laser frequency $\omega$.
Hence, the total energy $E_i$ of electron $i$ is approximately
\bea
E_i &\approx{}& 
\frac{\vec p_i^2}{2} -\frac{3Q_t}{2R_t} + 
\frac{\Omega_t^2}{2} \left(\overline{\vec{r}}_i
+\frac{\vec{F}(t)}{\Omega_t^2}\right)^2 
+\sum_{j(\ne i)}\frac{1}{|\vec r_j {-} \vec r_i|} 
\nonumber\\
&\approx & 
\frac{\vec p_i^2}{2} +V_t(r_i) + 
\overline{\vec{r}}_i\vec{F}(t) 
+\sum_{j(\ne i)}\frac{1}{|\vec r_j {-} \vec r_i|}\,.
\label{total}
\eea
To arrive at \eq{total} from \eq{eq:teei} we, firstly, replaced
the ionic potential by the cluster 
potential (\ref{eq:clpo}) and, secondly, neglected the
small term quadratic in $\vec{F}$.
From \eq{total} it becomes clear that by monitoring $E_i$ we
see essentially the coupling to the field, i.e., 
\be{sepctrum}
E_i = E_t^{(i)} + \overline{\vec{r}}_i \vec F(t)
\ee
where $E_t^{(i)}$ is the energy which varies slowly in time.
The spread in $E_i$ results from the different positions 
$\overline{\vec{r}}_i$ of the electrons within the cluster,  
$|\overline{\vec{r}}_i|\le R_t$. 
Indeed, the
spread is amplified through the laser field amplitude for maximum
field while the electron energy is focused to a small range at zero
field, in accordance with the observation in \fig{spectra_A50}
or more clearly in \fig{cut}.
 
\section{Summary}

We have analyzed the time-dependent distribution of
electron energies resulting from the 
illumination of  a large cluster containing about $10^4$ Xenon atoms
with a short intense laser pulse at 780\,nm wavelength. The
spectrum shows a rich structure on the time scale of the optical laser
period with many interesting details which would be worthwhile to
probe in the future using attosecond techniques. This might be even
possible in the near future since  
extended systems like a cluster show interesting electron dynamics on
the 100 attosecond time scale while more tightly bound electrons in
atoms or ions are too fast for direct probing with pulses of 10--100
attosecond length. 

Moreover, for now we do not have detailed knowledge about time
resolved electron motion in clusters and such experiments would also
reveal if our way of describing this dynamics theoretically is
adequate.


\end{document}